\documentclass[aip,jcp,preprint]{revtex4-1}
\usepackage{graphicx}

\usepackage{amsfonts, amsmath, amssymb,latexsym}

\usepackage{dcolumn}

\usepackage{bm}

\begin{document}

\title{Microstructures of capped ethylene oxide oligomers in water and $n$-hexane}

\author{Mangesh I. Chaudhari and Lawrence R. Pratt} 
\affiliation{Department of Chemical and Biomolecular Engineering, Tulane University, New Orleans, LA 70118}

\date{\today}

\begin{abstract} 
This report documents microstructural features of
CH$_3$(CH$_2$-O-CH$_2$)$_m$CH$_3$ dissolved in water and $n$-hexane for
$m$ = 11, 21, and 31.  Probability densities for end-to-end distance, and the
associated potential-of-mean-force (pmf), are more revealing of chain
microstructures than are the corresponding results for the radii of
gyration.  For water, the pmf identifies three distinct regions: 
loop-closure, globule, and high-extension regions.   The globule region
affirms a water-swollen chain, and is not evident in the
$n$-hexane results.  Chain C-atom density profiles from the chain
centroid are also different in the water and $n$-hexane cases.  
For $n$-hexane (but not water), the density profiles are similar for the
different chain lengths when the distances are scaled by the observed
$\left\langle R_\mathrm{g}{}^2\right\rangle^{1/2}$. For water (but not
$n$-hexane) and the smaller chains considered, the carbon material
exhibits a distinctive enhanced concentration, or internal condensation,
at the centroid core of the structure.
\end{abstract}

\maketitle 

\section{Introduction}

Dispersant materials used in response to oil spills\cite{dispersants}
undergo a demanding effectiveness approval process, and approved
dispersants are stockpiled near drilling activity and possible spill
locations. These factors inhibit continuous improvement of dispersants. Rigorous laboratory studies of the molecular-thermodynamic
interactions at play in dispersant systems are, therefore, particularly
welcome.  In addition, effective thermophysical modeling for these systems should
improve the fidelity of laboratory-scale  simulations of field-scale and ocean-scale
oil spill dispersion.  A typical prerequisite for development of
effective molecularly specific thermophysical models is computational simulation of
these solutions on a molecular basis.

Arguably the most important water-soluble synthetic
polymers,\cite{Alessi:2005ix,Norman:2007kq} (-CH$_2$-CH$_2$-O-)$_n$
chain molecules are also soluble in common organic solvents and are
intrinsic to the dispersant materials in use.\cite{dispersants}  This report establishes molecular-scale
microstructures of PEO oligomers in aqueous solutions, information
preparatory to molecular-scale theories of the structure and function of
these species in solution.

The molecular versatility of PEO/PEG polymers in solution makes them  a
challenge for molecularly specific statistical thermodynamic theory. For
example, their conformations respond sensitively to the solution
environment. \cite{Alessi:2005ix,Norman:2007kq} Common-sized PEO
molecules are distinctly helical in $n$-propanoic, isobutyric, and
isopentanoic acid solutions coexisting with liquid
water.\cite{Norman:2007kq} That helicity emphasizes the contrasting behavior
that these chains exhibit generic coil structures in
aqueous solutions.  Though the helices are not evident in bulk liquid
water, and indeed also not in acetic acid nor in isobutanol nor in
$n$-butanol, helix formation does seem to require at least a trace of
water.\cite{Norman:2007kq}

The solvent plays an intrinsic role in the solution thermodynamics also.
 The conformational sensitivity noted above is associated with a polymer
size fractionation between those coexisting
solutions.\cite{Alessi:2005ix,Norman:2007kq}   Experimental evaluation
of a Flory-Huggins interaction parameter for PEO/PEG in either water or
methanol shows  compositional dependence that is substantially different
in these two cases.\cite{Bae:1993uj,ZafaraniMoattar:2006kr}

In many applications,\cite{Lin:2012dz} and specifically for dispersants
used on oil spills,\cite{dispersants} the PEO/PEG chains
are decorated with junction, head, or capping groups. The work below
anticipates those molecular designs, but treats explicitly only the
simplest capping group, (methyl) -CH$_3$.  Correlations
associated with capping groups serve to focus  structural
analyses and the following work focuses on the
radial distributions of capping groups. Alternatively, spatial extensions
of chain molecules in solution are often characterized by a mean square-radius 
of gyration $\left\langle R_g{}^2\right\rangle$, and that information is addressed below for the
cases treated.  A conclusion we draw from the recent
exhaustive treatment of longer alkanes in water\cite{Ferguson:2009cf,Ferguson:2010kk}
is that distributions of $R_g{}^2$ are less informative of molecular structure than
spatial correlations between chemical groups such as capping groups. 
The results below should provide further clarification of that issue. 
We note that single-molecule pulling
experiments\cite{oesterhelt_single_1999,liu_extracting_2011,Gunari:2007cs,Li:2010bh,Li:2011ka,Li:2012jc}  
are becoming a new source of microstructural information, though not yet
at the molecular resolution of interest in the present work.  Those
important experiments do not track the radius of gyration
specifically, but instead manipulate lengths between tethered chemical groups.

Numerous  traditional statistical thermodynamics theories of aqueous
PEO/PEG solutions have been suggested; some representatives were noted
in our initial report,\cite{chaudhari_communication:_2010} and we do not
attempt a review here.  An alternative non-traditional procedure for
theory development accepts that simulations are necessary for evaluation
of molecular theories of liquids, that simulations will always be
carried-out in any case, and so the simulation work might as well
contribute to formulation of physical theories.   Of course, analyses
for those purposes must express the natural physical concepts
under-pinning molecular theories, and those analyses must be
sufficiently general.  We have elsewhere argued\cite{Chempath:2009ut}
that the recent developed molecular quasi-chemical theory does achieve
those requirements.   That molecular quasi-chemical theory has lead over
that past decade to resolution of the most basic puzzles of hydrophobic
effects.\cite{PrattLR:Molthe,AshbaughHS:Colspt} It is our intention to
establish, for the interesting and important aqueous solutions
considered here, some of the groundwork required for fully defensible
molecular theory.

As noted already, direct high-resolution molecular simulations are
necessary for the theory development that is sought.   Because these
systems are of practical importance, a wide range of simulation
calculations have been carried-out.   Again, we do not attempt a review
here, but merely note a recent contribution\cite{Starovoytov:2011km}
that gives access to that substantial available literature.

The results below were obtained by combination of standard molecular
simulation procedures appropriate for the sampling requirements of
different configurational aspects of CH$_3$(CH$_2$-O-CH$_2$)$_m$CH$_3$
chain molecules in water and $n$-hexane
solutions.\cite{chaudhari_communication:_2010} Further specific details
are provided in the Appendix.   We note here that the combined results
support comparisons over a range of molecule extensions, between
different chain molecule lengths, between different-sized simulation
systems, and between water and $n$-hexane solvents.

\begin{figure}[h]
\begin{center}
\includegraphics[width=7.0in]{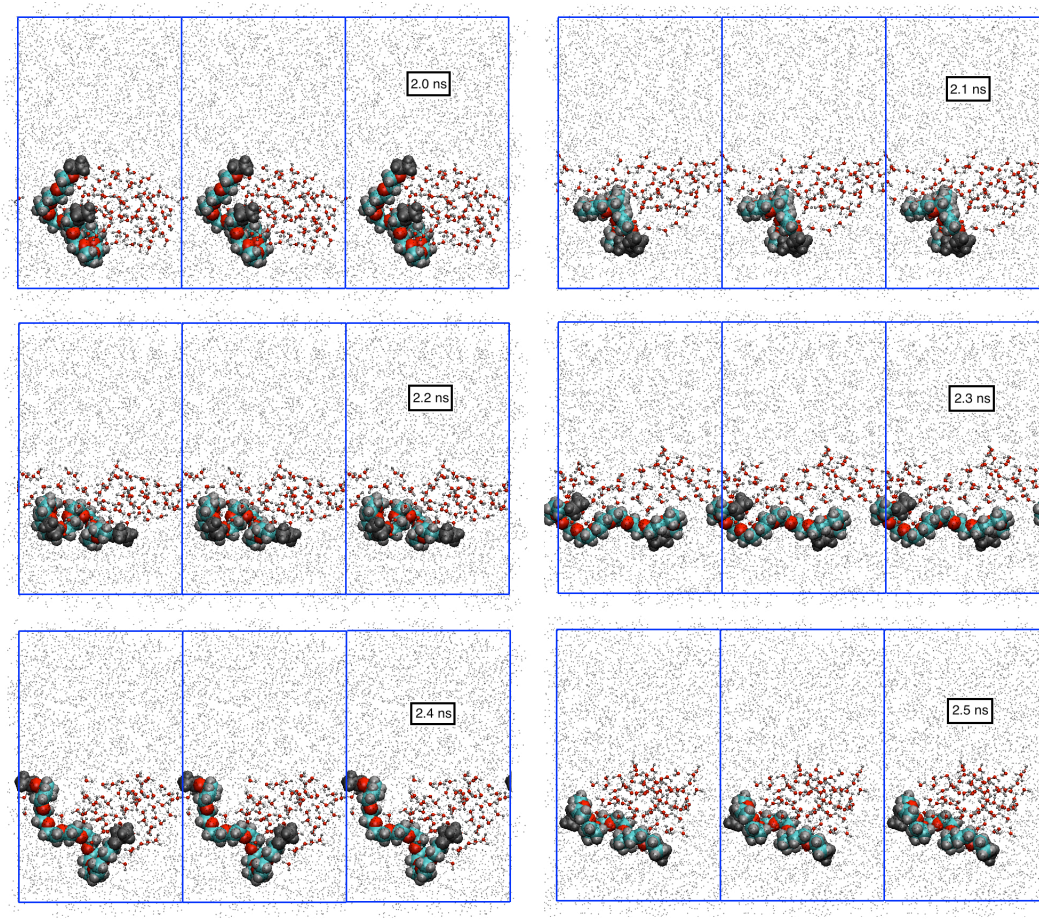}
\end{center}
\caption{Formation and then break-up of a fluid column by reconnection
across periodic boundary conditions.  For visual convenience, three
images are shown laterally in each frame. In this case the
droplet/column structure is composed of water with one
CH$_3$(CH$_2$-O-CH$_2$)$_m$CH$_3$ ($m$=11) chain molecule (in VDW
representation) that here resides near the interface. The majority
material in the background is $n$-hexane.\label{fig:fig1}}
\end{figure}

A curiosity of this simulation work is that with the typical time-scales and
length-scales  it is easy to observe 
(Fig.~\ref{fig:fig1}) molecular-scale break-up of a fluid column to
produce droplets, and then reformation of a fluid column.  We hope in
subsequent work to return to these observations with further analysis.

\section{Results and Discussion}

\begin{figure*}[h] 
\includegraphics[width=4in]{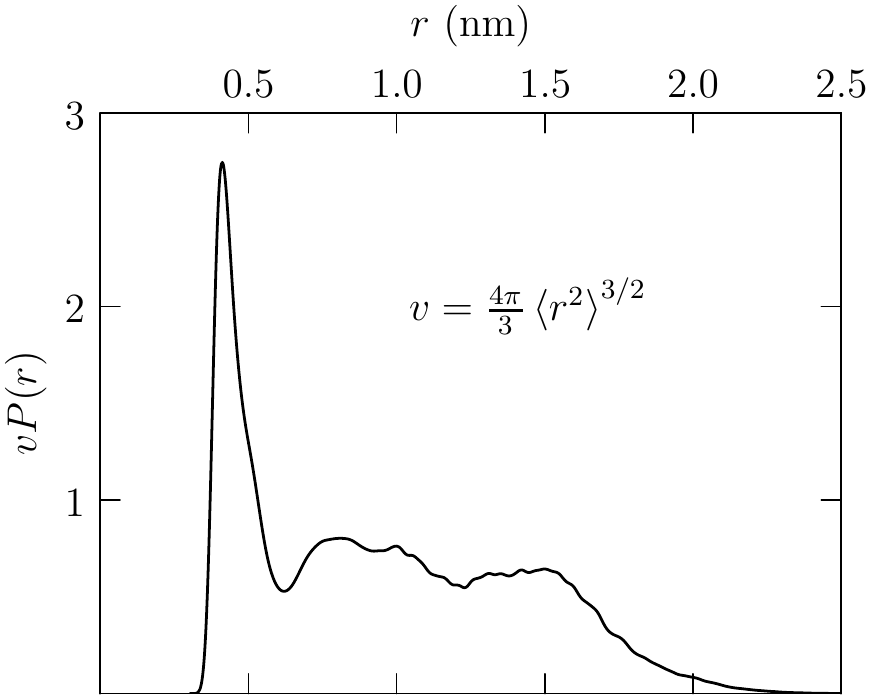}
\includegraphics[width=4in]{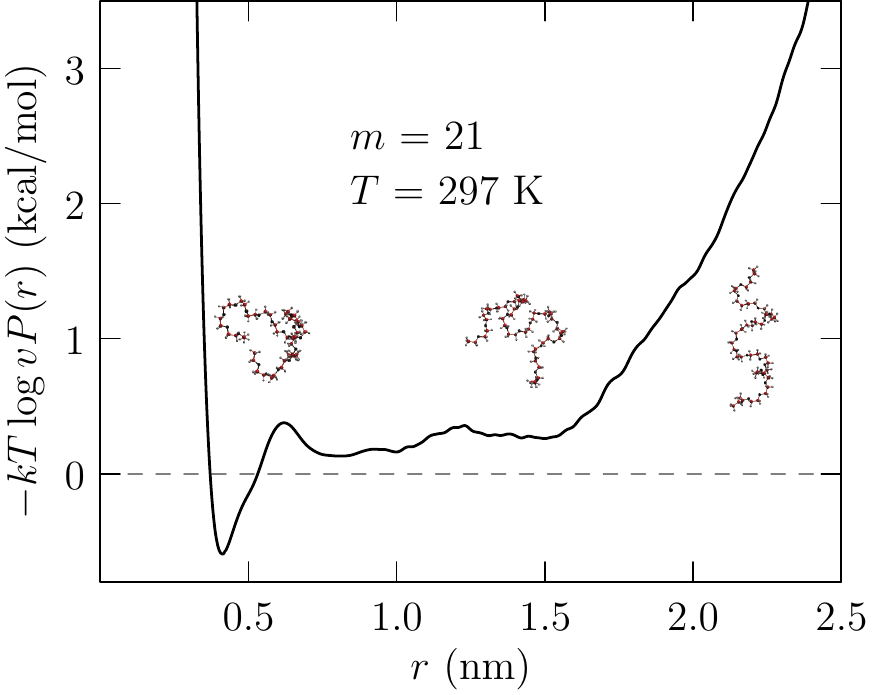}
\caption{(upper) Probability density $P(r)$ for end-to-end
(methyl-methyl) length for [CH$_3$(CH$_2$-O-CH$_2$)$_m$CH$_3$](aq) with
$m$ = 21.   The normalization of this graph is chosen to suggest the analogy with
conventional atom-atom radial distribution function in liquids. (lower)
Potential of the average end-to-end forces  showing distinct loop-closure,
globule, and high-extension regions. The choice of normalization for the
upper panel sets the origin of the y-axis of the lower panel, and the
same convention is followed in both cases.  Results here for $r <
1.0$~nm were  are obtained utilizing the WHAM procedure. Those high
resolution results were matched to overall observation of $P(r)$ from
molecular simulations with parallel tempering.
\label{fig:fig2}}
\end{figure*}

\begin{figure}[h]
\includegraphics[width=4.0in]{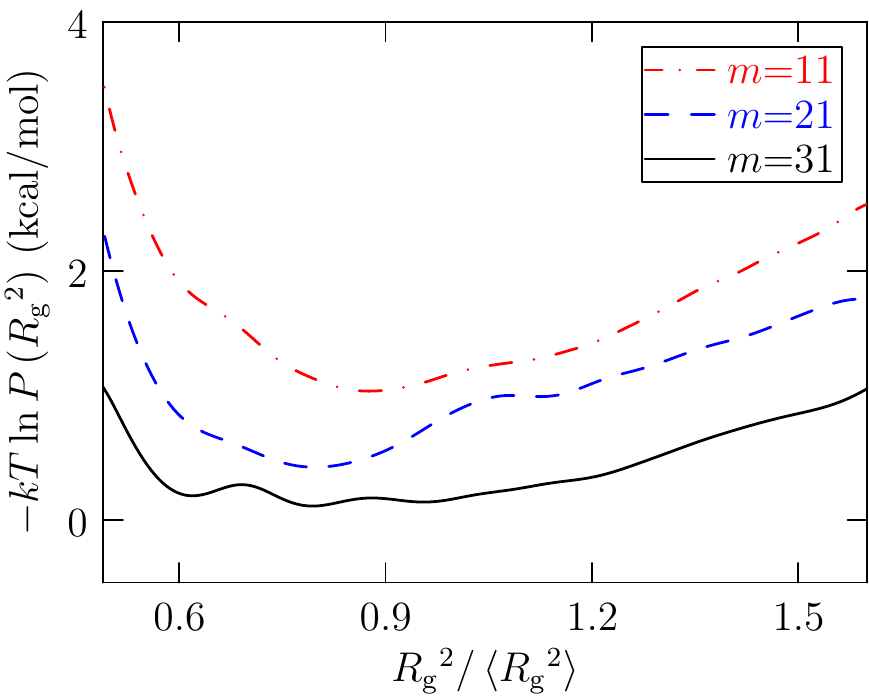}
\caption{For [CH$_3$(CH$_2$-O-CH$_2$)$_m$CH$_3$](aq), distributions of
square-radius of gyration displaced vertically for visual convenience,
extracted utilizing GROMACS analysis tools.   Scaling of the abscissa
brings the results for different lengths into a similar range:
$\left\langle R_\mathrm{g}{}^2\right\rangle^{1/2}$ = 0.61, 0.84, and
1.22~nm for $m$ = 11, 21, and 31, respectively. These values are
larger than the ideal $\left(1/\sqrt{6}\right)$ proportion of the observed
$\left\langle r^2\right\rangle^{1/2}$
\label{fig:fig3}}
\end{figure}

From small to large end-to-end lengths, the observed probability density
(Fig.~\ref{fig:fig2}) displays three distinct behaviors, loop-closure,
globule, and high-extension (elastic)
regions.\cite{chaudhari_communication:_2010}  The loop-closure feature
is a realization of  a primitive hydrophobic
bond,\cite{chaudhari_communication:_2010} which has been extensively
studied by simulation but not yet susceptible to experimental
measurement due to solubility limitations.\cite{Asthagiri:2008in} In
contrast (Fig.~\ref{fig:fig3}), distributions of $R_\mathrm{g}{}^2$ are
less distinctive of these microstructural features.

\begin{figure}[h]
\includegraphics[width=4.0in]{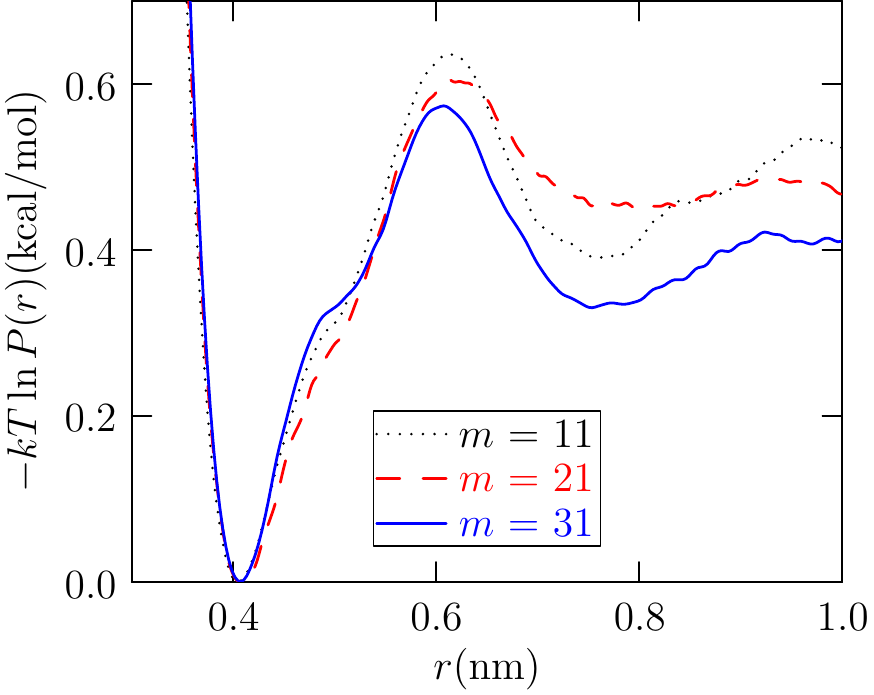}
\caption{High resolution results for $ r <$ 1~nm for several chain
lengths from windowing/WHAM calculations.\label{fig:fig4}}
\end{figure}

\begin{figure}[h]
\includegraphics[width=4.0in]{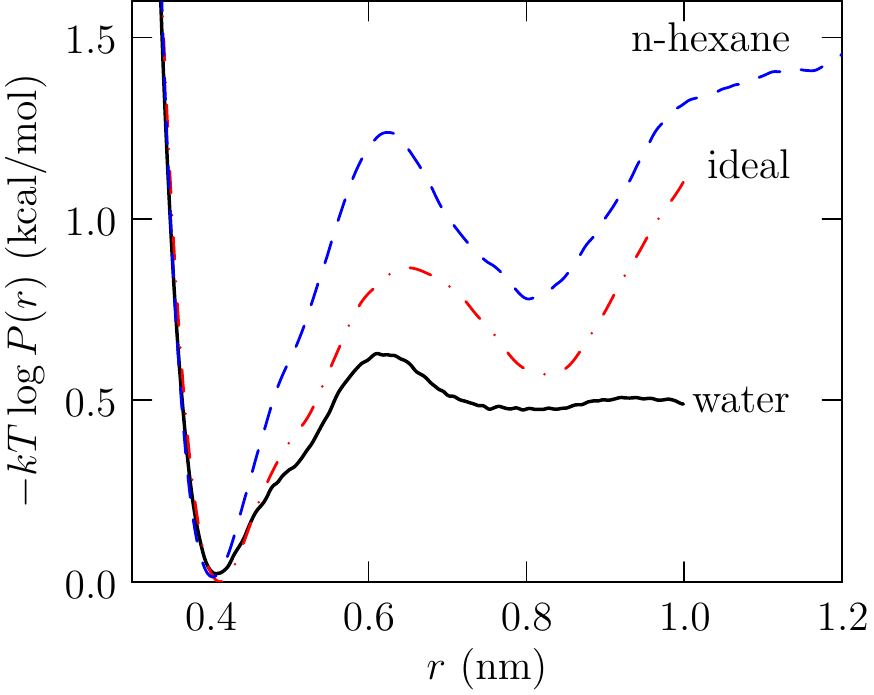}
\caption{In the loop-closure region, the [CH$_3$(CH$_2$-O-CH$_2$)$_m$CH$_3$]
($m$ = 21) oligomer behaves differently in $n$-hexane and water,
and the results for $n$-hexane are similar to those obtained with
no solvent (ideal gas case). The temperature variation here
is small, $T$ = 297~K (water), 302~K (ideal), and 297~K ($n$-hexane).
\label{fig:fig5}}
\end{figure}

For liquid water the loop-closure feature is largely independent of
oligomer size (Fig.~\ref{fig:fig4}). This feature in liquid water is
different from what is observed in $n$-hexane and for non-solvated
chains (Fig.~\ref{fig:fig5}).   These points support the
identification of the loop-closure feature as an experimentally
realizable hydrophobic bond.

\begin{figure}[h]
\includegraphics[width=4.0in]{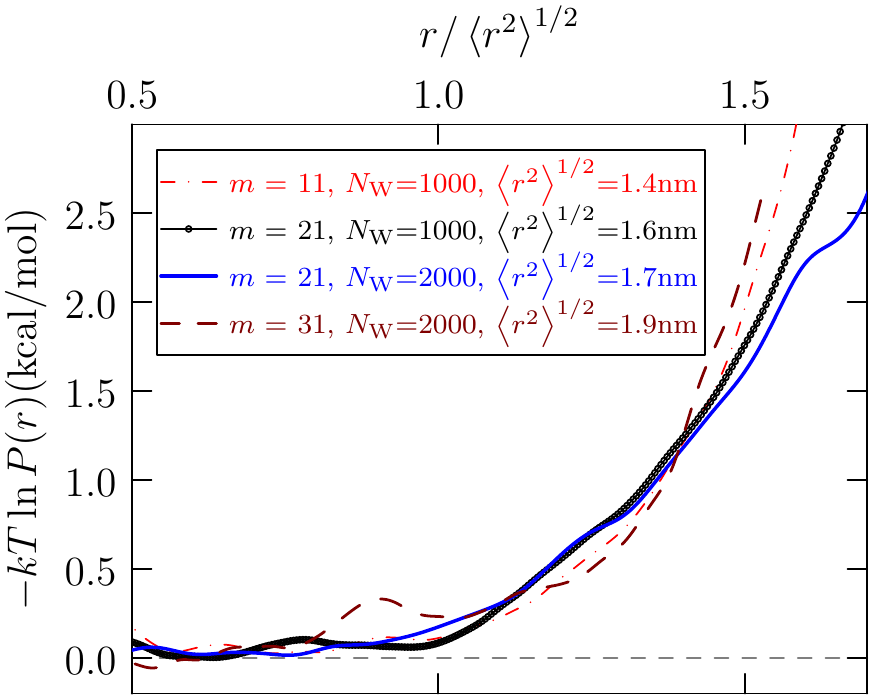}
\includegraphics[width=4.0in]{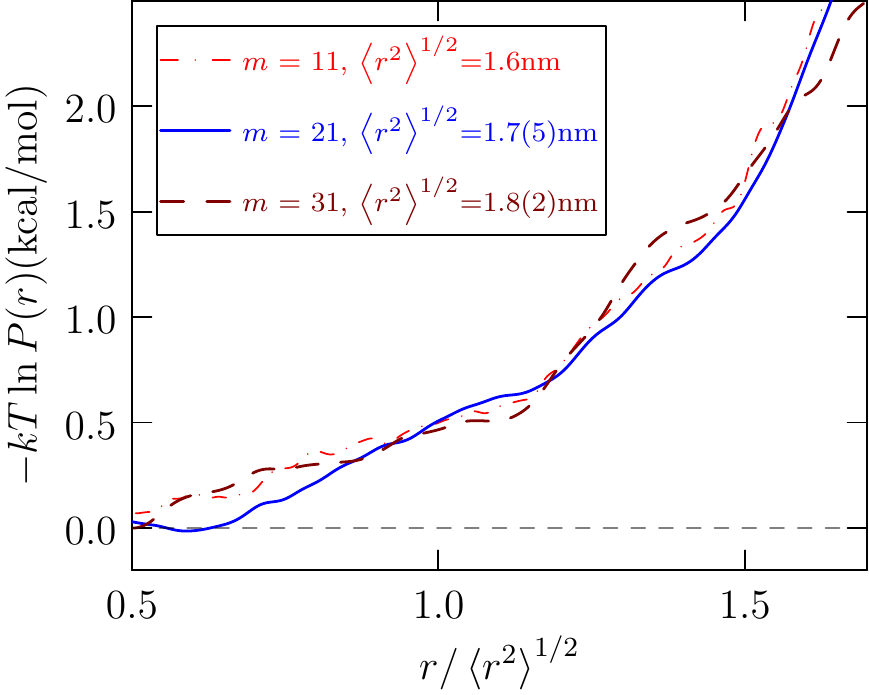}
\caption{(upper) For CH$_3$(CH$_2$-O-CH$_2$)$_m$CH$_3$ (aq) chains, 
at constant density corresponding to 300K and $p$ = 1 atm: parallel
tempering results for different chain lengths and system sizes overlap
each other on normalized end-to-end distance. Globule and high-extension
regions are separated at $r \approx \left\langle r^2\right\rangle^{1/2}$
for these chain lengths and system sizes. (lower) Corresponding results
for $n$-hexane at constant  density ($T$= 300K and $p$ = 1 atm):
Separation of a globule region and a high extension region is indistinct
here. A Gaussian (parabola) model
satisfactorily fits the data in this high extension
regime.\label{fig:fig6}}
\end{figure}

The globule region is clear for the hydrated chains, but not for the
case of $n$-hexane solvent (Fig.~\ref{fig:fig6}).  Thus in the more
benign organic solvent the chains are collapsed relative to the more
swollen hydrated oligomers. This also supports the interpretion of
intrachain interactions from the perspective of simple aqueous solution
examples.

\begin{figure*}[h] 
\includegraphics[width=4in]{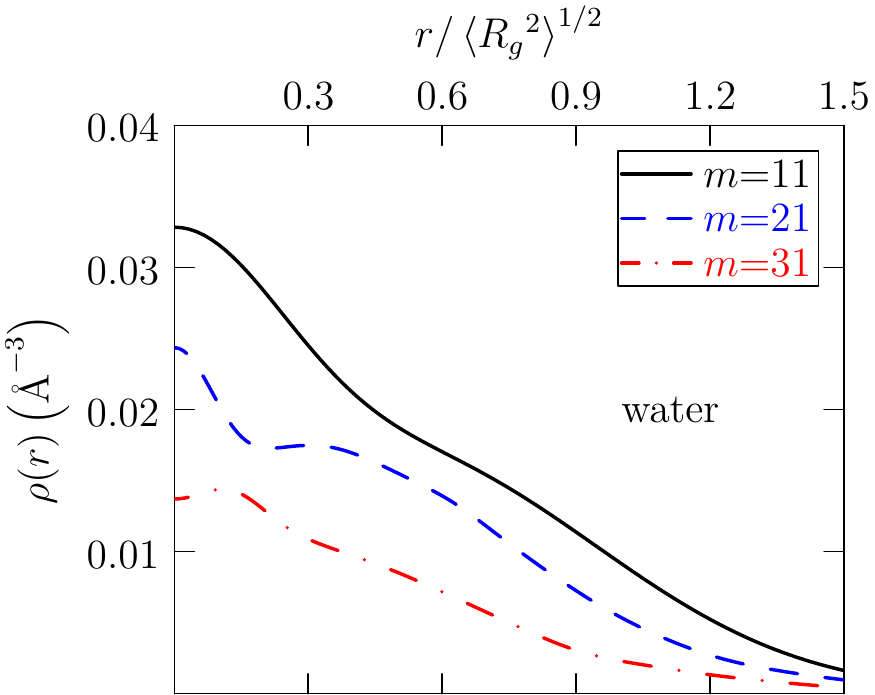}
\includegraphics[width=4in]{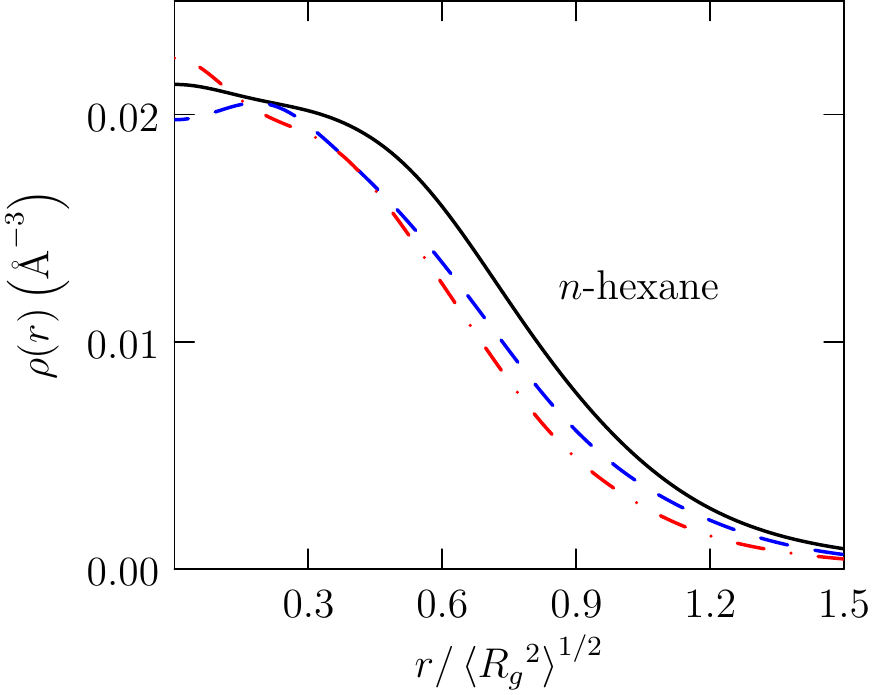}
\caption{CH$_3$(CH$_2$-O-CH$_2$)$_m$CH$_3$ chain C-atom density profiles
from the chain centroid shows higher concentration of polymer C-atoms in
water the middle of distribution which is absent for $n$-hexane
solvated polymer\label{fig:fig7}}
\end{figure*}

The  profiles of the density of chain C-atoms, measured relative to the
chain centroid, are different
(Fig.~\ref{fig:fig7}) for the two solvents.  For $n$-hexane,
the density profiles are similar for the different chain lengths when
the distances are scaled by the observed $\left\langle
R_\mathrm{g}{}^2\right\rangle^{1/2}$. This similarity does not obtain
for the case for water.  In water and for the smaller chains studied, the
carbon material exhibits a distinctive enhanced concentration in the
middle of the distribution, a concentration that is not evident for the
case of $n$-hexane.   This interesting internal condensation
is characterized by a length-scale smaller  than $\left\langle
R_\mathrm{g}{}^2\right\rangle^{1/2}$.  Thus the overall
density profiles show less similarity on the $\left\langle
R_\mathrm{g}{}^2\right\rangle^{1/2}$ scale, and the internal
condensation feature is not evident for the largest $m$=31 chain.

\section{Conclusions} 
The probability density for end-to-end distance (and the
associated pmf) for CH$_3$(CH$_2$-O-CH$_2$)$_m$CH$_3$ is
more revealing of chain microstructures than are the corresponding
results for the radii of gyration.  For water, the pmf identifies three
distinct regions: loop-closure, globule, and high-extension regions. In
water, the loop-closure feature is similar to a primitive hydrophobic
bond and is insensitive to chain length. The globule region exposes
a water-swollen chain, and is not evident in the $n$-hexane results. 
Chain C-atom density profiles from the chain centroid are
different in the water and $n$-hexane cases. For $n$-hexane (but not
water), the density profiles are similar for the different chain lengths
when the distances are scaled by the observed $\left\langle
R_\mathrm{g}{}^2\right\rangle^{1/2}$. For water (but not $n$-hexane) and
the smaller chains considered, the carbon material exhibits a
distinctive, enhanced concentration, or internal condensation, at the
centroid core of the structure.

\appendix*
\section{Simulation Specifics}
Two different simulation techniques were used for the molecular dynamics
calculations. The first class of calculations used parallel
tempering\cite{Earl:2005fv} to achieve enhanced sampling of the mixing
characteristics with water or $n$-hexane solvent, and of
CH$_3$(CH$_2$-O-CH$_2$)$_m$CH$_3$ chain conformations. The chain
molecules were represented by a generalized amber force field
(GAFF),\cite{Wang:2004fe} the SPC/E  model for
water,\cite{Berendsen:1987uu} and optimized potentials for liquid
simulations (OPLS-AA) were used to describe $n$-hexane in those
simulations.\cite{Jorgensen:1996vx} GROMACS 4.5.3 molecular dynamic
simulation package\cite{vanderSpoel:2005hz}  was used for all parallel
tempering simulations spanning the 256-550K temperature range  with 32
replicas (for $m$ = 11, and 21, cases) and 40 replicas (for $m$ = 31).
Long-range electrostatic interactions were treated in standard periodic
boundary conditions using the particle mesh Ewald method with a cutoff
of 0.9nm. The Nos\'{e}-Hoover thermostat maintained the constant temperature
and chemical bonds involving hydrogen atoms were constrained by the
LINCS algorithm. Energy minimization and constant pressure density
equilibration was performed at 300K to set the constant volume
conditions, and then production calculations for each replica set were
extended to 10 ns.   Parallel tempering swaps were attempted at a rate
of 100/ns, and the temperature grid chosen resulted in a success rates of
15-25\%.

The second class of simulations obtained higher spatial resolution in
the loop closure region using the windowing stratification
method.\cite{Shell,AlanGrossfield} In these simulations a harmonic
interaction between the  capping atoms of the
CH$_3$(CH$_2$-O-CH$_2$)$_m$CH$_3$ chain  was exploited  to concentrate
the sampling of the end-to-end distance to each particular window. For a
CH$_3$(CH$_2$-O-CH$_2$)$_m$CH$_3$ chain in water or $n$-hexane, we
performed simulations with radial displacement coordinate covering the
range from 3.0-10.0{\AA} uniformly with 15 windows. Trajectories were
recorded for 10ns/window at the temperature of 297.1K. $P(r)$ was then
reconstructed over the whole range using the weighted histogram analysis
method (WHAM).\cite{Shell,AlanGrossfield}

\clearpage



\providecommand*\mcitethebibliography{\thebibliography}
\csname @ifundefined\endcsname{endmcitethebibliography}
  {\let\endmcitethebibliography\endthebibliography}{}

\end{document}